\documentclass[preprint,showpacs]{revtex4-1}

\usepackage{graphicx}
\usepackage{epsfig}
\usepackage{dcolumn}
\usepackage{bm}

\def\ktf {$k_T$-factorization }
\def\ktfa {$k_T$-factorization approach }
\def\cpc#1#2#3  {{Computer\ Phys.\ Comm.}  {\bf#1}, #2 (#3)}
\def\err#1#2#3  {{\it Erratum }              {\bf#1}, #2 (#3)}
\def\epjc#1#2#3 {{Eur. Phys. J. C}          {\bf#1}, #2 (#3)}
\def\dum#1#2#3  {{~}                         {\bf#1}, #2 (#3)}
\def\ib#1#2#3   {{\it ibid. }                {\bf#1}, #2 (#3)}
\def\jcp#1#2#3  {{J.\ Comput.\ Phys.\ }      {\bf#1}, #2 (#3)}
\def\jetpl#1#2#3 {{\rm JETP Lett.}           {\bf#1}, #2 (#3)}
\def\jhep#1#2#3 {{JHEP }                     {\bf#1}, #2 (#3)}
\def\ijmp#1#2#3 {{Int.\ J.\ Mod.\ Phys.\ }   {\bf#1}, #2 (#3)}
\def\jpg#1#2#3  {{J.\ Phys.\ G }             {\bf#1}, #2 (#3)}
\def\mpl#1#2#3  {{Mod.\ Phys.\ Lett.\ }      {\bf#1}, #2 (#3)}
\def\mpla#1#2#3 {{Mod.\ Phys.\ Lett.\ A }    {\bf#1}, #2 (#3)}
\def\ncim#1#2#3 {{Nuovo Cimento }            {\bf#1}, #2 (#3)}
\def\np#1#2#3   {{Nucl.\ Phys.}            {\bf#1}, #2 (#3)}
\def\npb#1#2#3  {{Nucl.\ Phys. B}           {\bf#1}, #2 (#3)}
\def\pan#1#2#3  {{Phys.\ At.\ Nuclei }       {\bf#1}, #2 (#3)}
\def\plb#1#2#3  {{Phys.\ Lett. B}          {\bf#1}, #2 (#3)}
\def\prep#1#2#3 {{Phys.\ Rep.}             {\bf#1}, #2 (#3)}
\def\prd#1#2#3  {{Phys.\ Rev. D}           {\bf#1}, #2 (#3)}
\def\prl#1#2#3  {{Phys.\ Rev.\ Lett.\ }      {\bf#1}, #2 (#3)}
\def\ptp#1#2#3  {{Prog.\ Theor.\ Phys.\ }    {\bf#1}, #2 (#3)}
\def\ps#1#2#3   {{Physica Scripta }          {\bf#1}, #2 (#3)}
\def\rmp#1#2#3  {{Rev.\ Mod.\ Phys.\ }       {\bf#1}, #2 (#3)}
\def\rpp#1#2#3  {{Rep.\ Prog.\ Phys.\ }      {\bf#1}, #2 (#3)}
\def\sa#1#2#3   {{Sci. Acta}                 {\bf#1}, #2 (#3)}
\def\sjnp#1#2#3 {{Sov.\ J.\ Nucl.\ Phys.\ }  {\bf#1}, #2 (#3)}
\def\spj#1#2#3  {{Sov.\ Phys.\ JETP }        {\bf#1}, #2 (#3)}
\def\spjl#1#2#3 {{Sov.\ JETP Lett.\ }        {\bf#1}, #2 (#3)}
\def\spu#1#2#3  {{Sov.\ Phys.-Usp.\ }        {\bf#1}, #2 (#3)}
\def\yaf#1#2#3  {{Yad.\ Fiz.\ }              {\bf#1}, #2 (#3)}
\def\zp#1#2#3   {{Zeit.\ Phys.\ }            {\bf#1}, #2 (#3)}
\def\zpc#1#2#3  {{Z.\ Phys.\ C }             {\bf#1}, #2 (#3)}

\begin{document}

\begin{flushright}
DESY 16-180
\end{flushright}

\title{Testing the parton evolution with the use of two-body final states}

\author{S.P.~Baranov} 
\affiliation{P.N.~Lebedev Physics Institute, 119991 Moscow, Russia}

\author{H.~Jung}
\affiliation{Deutsches Elektronen-Synchrotron, Notkestrasse 85, Hamburg, Germany}

\author{A.V.~Lipatov}
\affiliation{Skobeltsyn Institute of Nuclear Physics, Lomonosov Moscow State University, 119991 Moscow, Russia}
\affiliation{Joint Institute for Nuclear Research, Dubna 141980, Moscow region, Russia}

\author{M.A.~Malyshev}
\affiliation{Skobeltsyn Institute of Nuclear Physics, Lomonosov Moscow State University, 119991 Moscow, Russia}

\begin{abstract}

We consider the production of $b\bar b$ quarks and Drell-Yan lepton 
pairs at LHC conditions focusing attention on the total 
transverse momentum of the produced pair and on the azimuthal 
angle between the momenta of the outgoing particles.
Plotting the corresponding distributions in bins of the final state 
invariant mass, one can reconstruct the full map of the transverse momentum dependent 
parton densities in a proton. 
We give examples of how can these distributions can look like 
at the LHC energies. 

\end{abstract} 


\maketitle

\newpage 

Experiments of new generation running at the LHC yield plenty of 
high precision data. In order to properly interpret these data we
need that the parton distribution functions to be known with adequately 
good accuracy. This, in turn, rises question on a detailed measurement 
of parton distributions. 
In this note we focus attention on two important kinematic observables 
which enable us to reconstruct the full map of the transverse momentum dependent
(TMD), or unintegrated, parton densities. We address our consideration to 
the LHC conditions ($pp$ collisions at $\sqrt{s}=7$ TeV), for which 
we give a number of illustrations.

The evolution of TMD gluon densities can be explored with the production 
of $b\bar b$ pairs. At the LHC energies, this process is 
dominated by the direct leading-order (LO) off-shell gluon-gluon fusion subprocess
\begin{equation}\label{bb}
 g^*(k_1)+g^*(k_2) \to b(p_1) + \bar b(p_2), 
\end{equation}
\noindent 
while the contribution from the quark-antiquark annihilation
is of almost no importance because 
of comparatively low quark densities.
The four-momenta of corresponding particles are given in the parentheses.
The present calculation of the process (\ref{bb})
is fully identical to that performed previously~\cite{JKLZ}.
The evolution of TMD quark densities can be explored with
the production of Drell-Yan lepton pairs. This process is dominated
by the off-shell quark-antiquark ahhihilation subprocess
\begin{equation}\label{qq_DY}
q^*(k_1) + \bar q^*(k_2) \to l^+(p_1) + l^-(p_2),
\end{equation}
where $q$ includes valence and sea quarks and $\bar{q}$ stands for sea
anti-quarks. 
The present calculation of the process (\ref{qq_DY})
is fully identical to that from~\cite{BLZ_DY}.
We do not consider here higher-order corrections 
$q + \bar{q} \to l^+ + l^- + g$ since they are already taken into account 
in the \ktfa~\cite{GLR83,Catani,Collins} as a part of the evolution of TMD quark densities. 

The final states of the processes (\ref{bb}) and (\ref{qq_DY}) are represented by two-body 
systems with fully reconstructable kinematics where the transverse
momentum $p_T$ of the $b\bar b$ or lepton pair measures the net transverse momentum 
of the initial gluons or quarks, the invariant mass of the pair measures the 
product of longitudinal momentum fractions, $M^2 = x_1x_2s$, 
and the rapidity of the pairs measures the ratio of the momentum
fractions, $y = (1/2)\ln(x_1/x_2)$.
A useful complementary observable is the difference between the 
azimuthal angles of produced particles $\Delta \phi$.
In the LO of collinear QCD factorization, the $p_T$ and 
$\Delta \phi$ distributions degenerate into delta functions at $p_T = 0$ 
and $\Delta \phi = 0$, and the continuous spectra can only be obtained by
including higher-order corrections. In the \ktf approach, these 
radiative corrections are automatically taken into account in the 
form of TMD parton densities. Comparing the $p_T$ and
$\Delta \phi$ spectra at varying gluon momentum fraction $x$ we watch the
evolution of parton distributions.

To simulate the $b\bar b$ pair production we used the 
latest JH'2013 parametrization~\cite{JH}
for the TMD gluon densities in a proton.
The input parameters of this gluon distribution were
fitted to describe the proton structure function $F_2$.
To simulate the production of Drell-Yan lepton pairs we applied complementary
TMD valence quark distributions from the same set~\cite{JH}. The necessary TMD sea quark 
densities are calculated from the gluon ones in the
approximation where the sea quarks occur in the last gluon-to-quark splitting~\cite{HHJ}.

The results of our calculations are displayed in Figs.~1 --- 7.
Shown in Figs.~1 and 2 are the spectra of $b \bar b$ pair and dilepton 
transverse momentum $p_T$ and the azimuthal angle $\Delta \phi$
plotted for several different intervals of their invariant 
mass $M$. Here, to make the changes in shape easier recognizable,
we show the normalized differential cross sections.
We see that with increasing $M$ the maximum in the $p_T$ 
spectrum shifts gradually to higher values, and the whole distribution 
becomes more flat. The $\Delta \phi$ distribution moves towards 
$\Delta \phi \simeq \pi$, that is due to the inequality $M \gg p_T$.
The latter becomes even stronger at high $M$ (see Fig.~3).
As one can see from
Fig.~2, quark distributions follow the same trend as gluon densities.

The observed behaviour of calculated $p_T$ and $\Delta \phi$ distributions
is related to the different regions of
$x$ and/or parton transverse momenta probed in the considered $M$ bins.
In fact, with increasing of $M$, the achieved $x$ values
shifted towards unity, irrespectively on the 
rapidities of final-state particles, as it is 
demonstrated in Figs.~4 and 5.
The latter results to decreasing of average parton transverse
momentum generated in the non-collinear parton evolution.
At the highest $M$ bin, this average parton transverse
momentum becomes small compared to the hard scale (which is order of $M$), 
so that the collinear kinematics of the partonic subprocesses are reproduced.

Besides the restrictions on the invariant mass, the special kinematical cuts on the final state give us 
further possibilities to achive the wanted region of $x$ and/or partonic
transverse momenta. It is illustrated in Figs.~6 and 7, where 
we plot the normalized differential cross sections of the considered subprocesses 
calculated as a functions of $x$ and
${\mathbf k}_T^2$ (the longitudinal momentum fraction and transverse 
momentum of one of the colliding partons) with the additional
cuts applied to the rapidity $y$ of the final state quark or lepton pair.
As an example, we used $y < 1$ and $3 < y < 4$.
We show that under these cuts one can probe different $x$ and/or ${\mathbf k}_T^2$ regions 
and extract an information on the TMD parton distributions
at the scale given by $M$. 
Note that the different ${\mathbf k}_T^2$ regions 
can be achieved under additional restrictions on the quark or lepton pair 
transverse momentum $p_T$ and/or azimuthal angle $\Delta \phi$.

Thus, we conclude that 
one can map the evolution of parton distributions at the scale $M$ from high values 
of proton longitudinal momentum fraction $x$ to low ones
by applying different cuts on the final states.
This is important to further precise determination 
of the TMD quark and gluon densites in a proton from the LHC data.

\indent
{\sl Acknowledgements.} This research was supported in part by RFBR grant 16-32-00176-mol-a and
grant of the President of Russian Federation NS-7989.2016.2.
We are also grateful to DESY Directorate for the
support in the framework of Moscow---DESY project on Monte-Carlo implementation for
HERA---LHC.

\newpage

\begin{figure}
\begin{center}
\epsfig{figure=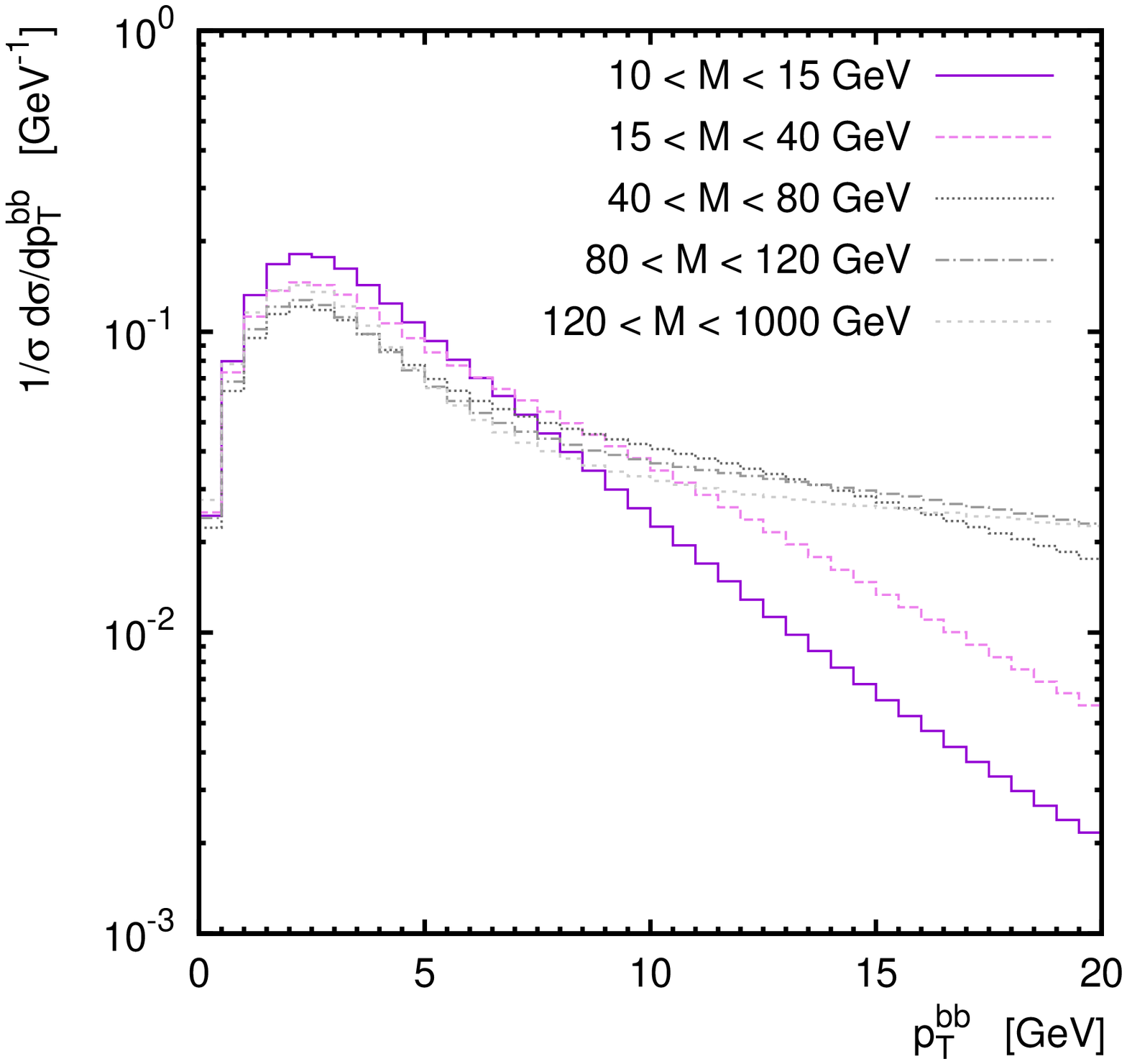, width = 6cm, angle = 270}
\hspace{-1cm}
\epsfig{figure=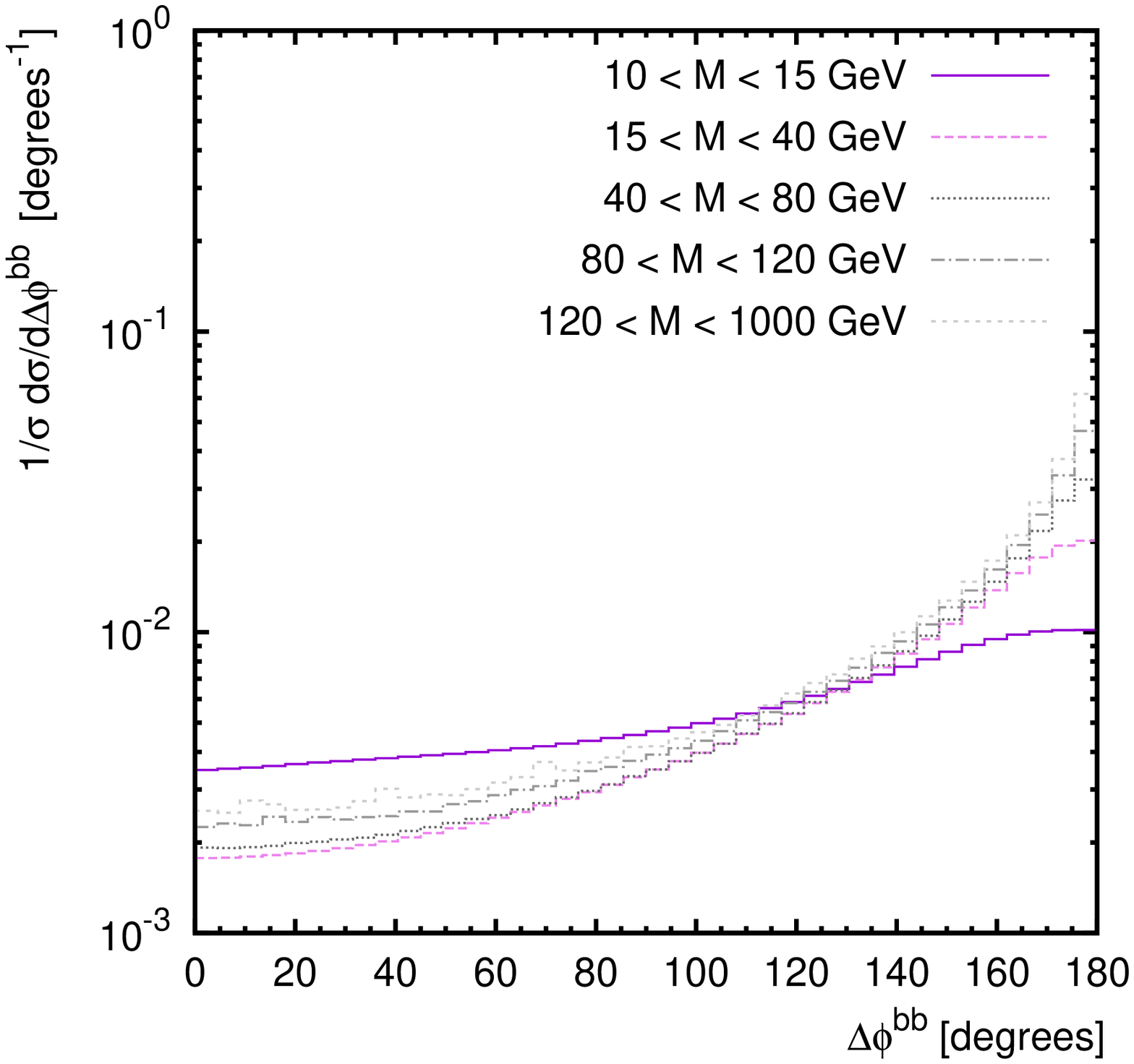, width = 6cm, angle = 270}
\caption{Spectra of the $b \bar b$ pair transverse momentum $p_T$
and the azimuthal angle between the beauty quarks $\Delta \phi$
for several different intervals of the $b\bar b$ invariant mass $M$.}
\label{fig1}
\end{center}
\end{figure}

\begin{figure}
\begin{center}
\epsfig{figure=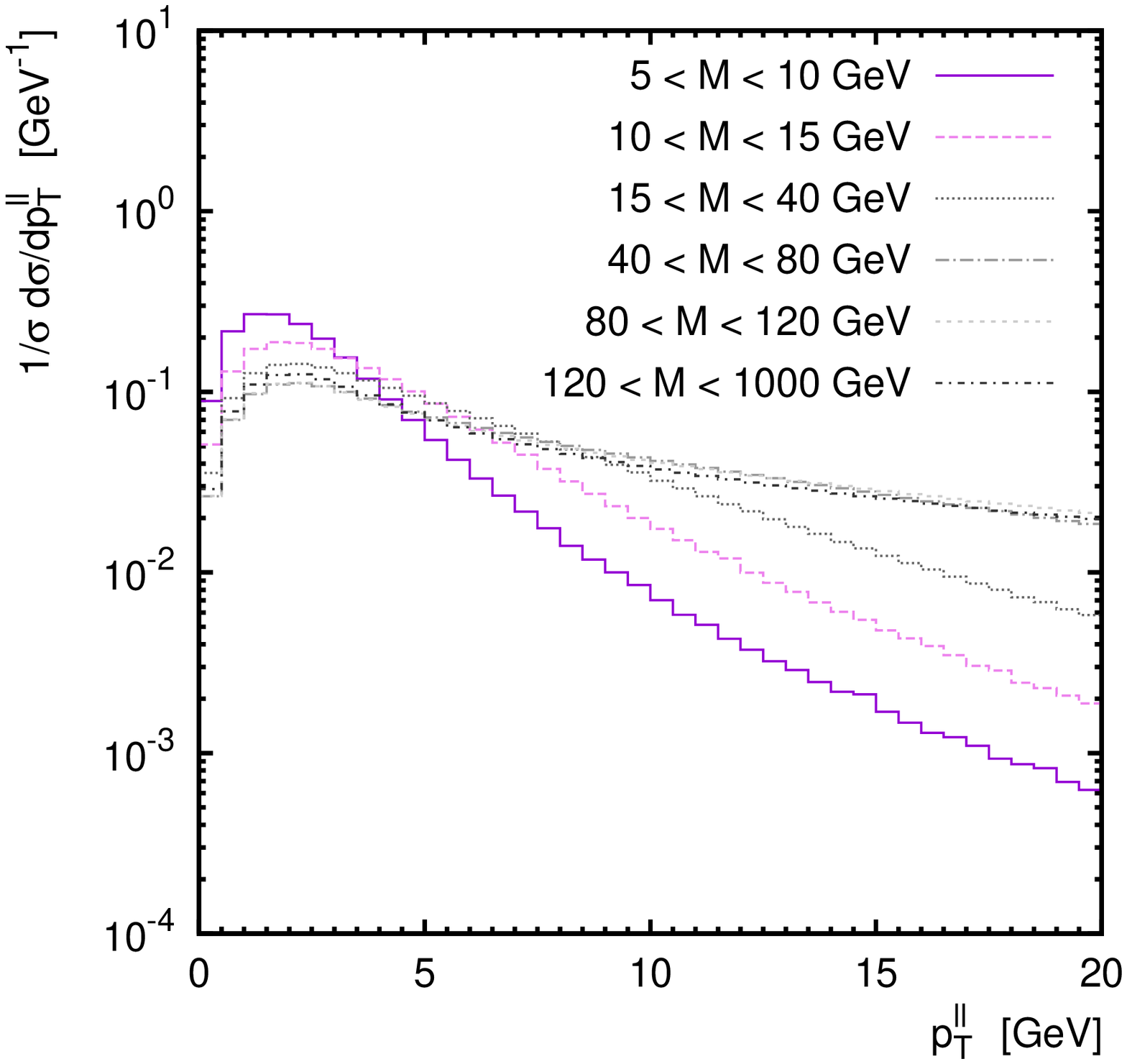, width = 6cm, angle = 270}
\hspace{-1cm}
\epsfig{figure=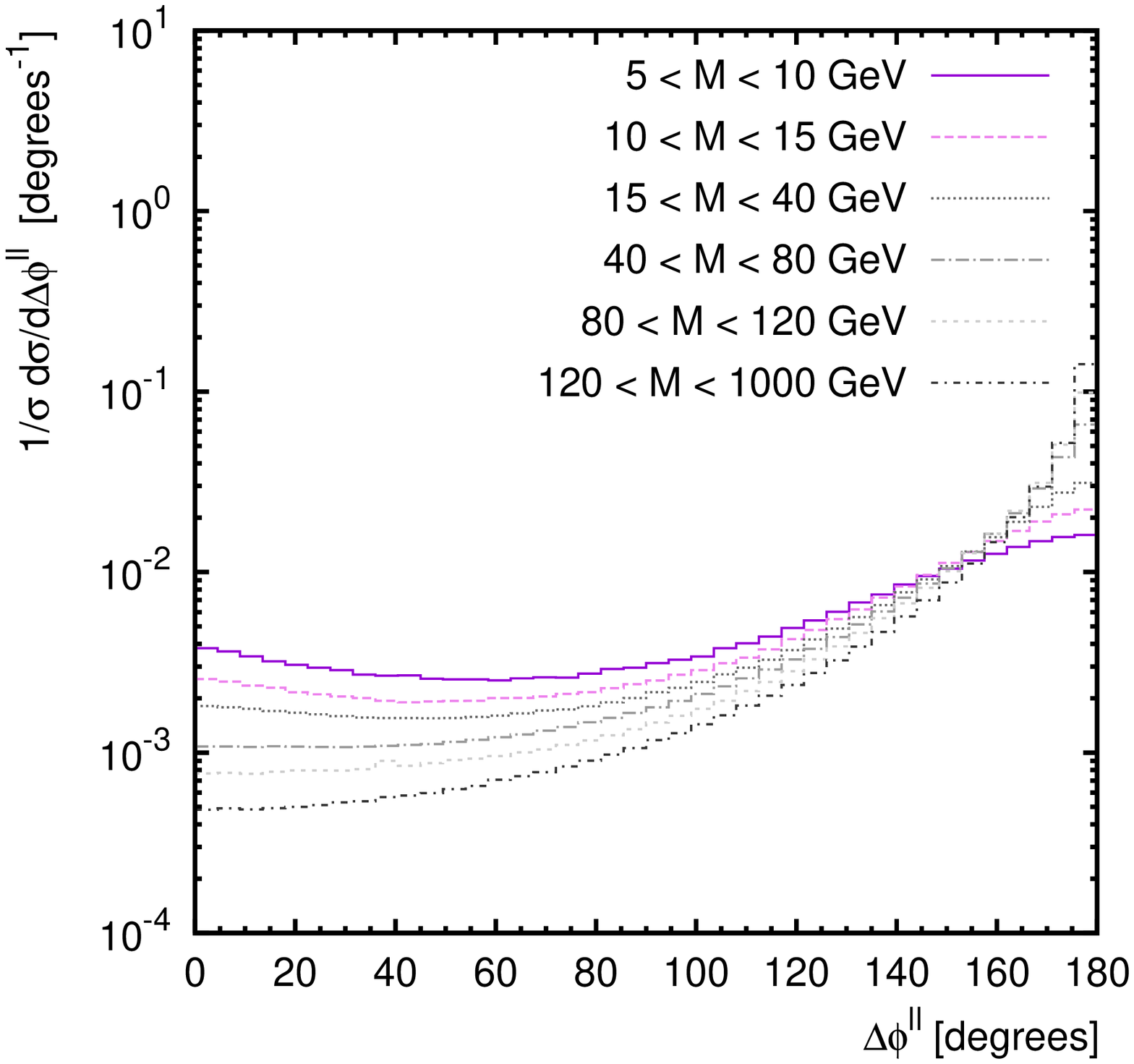, width = 6cm, angle = 270}
\caption{Spectra of the Drell-Yan lepton pair transverse momentum $p_T$
and the azimuthal angle between the produced leptons $\Delta \phi$
for several different intervals of the dilepton invariant mass $M$.}
\label{fig2}
\end{center}
\end{figure}

\begin{figure}
\begin{center}
\epsfig{figure=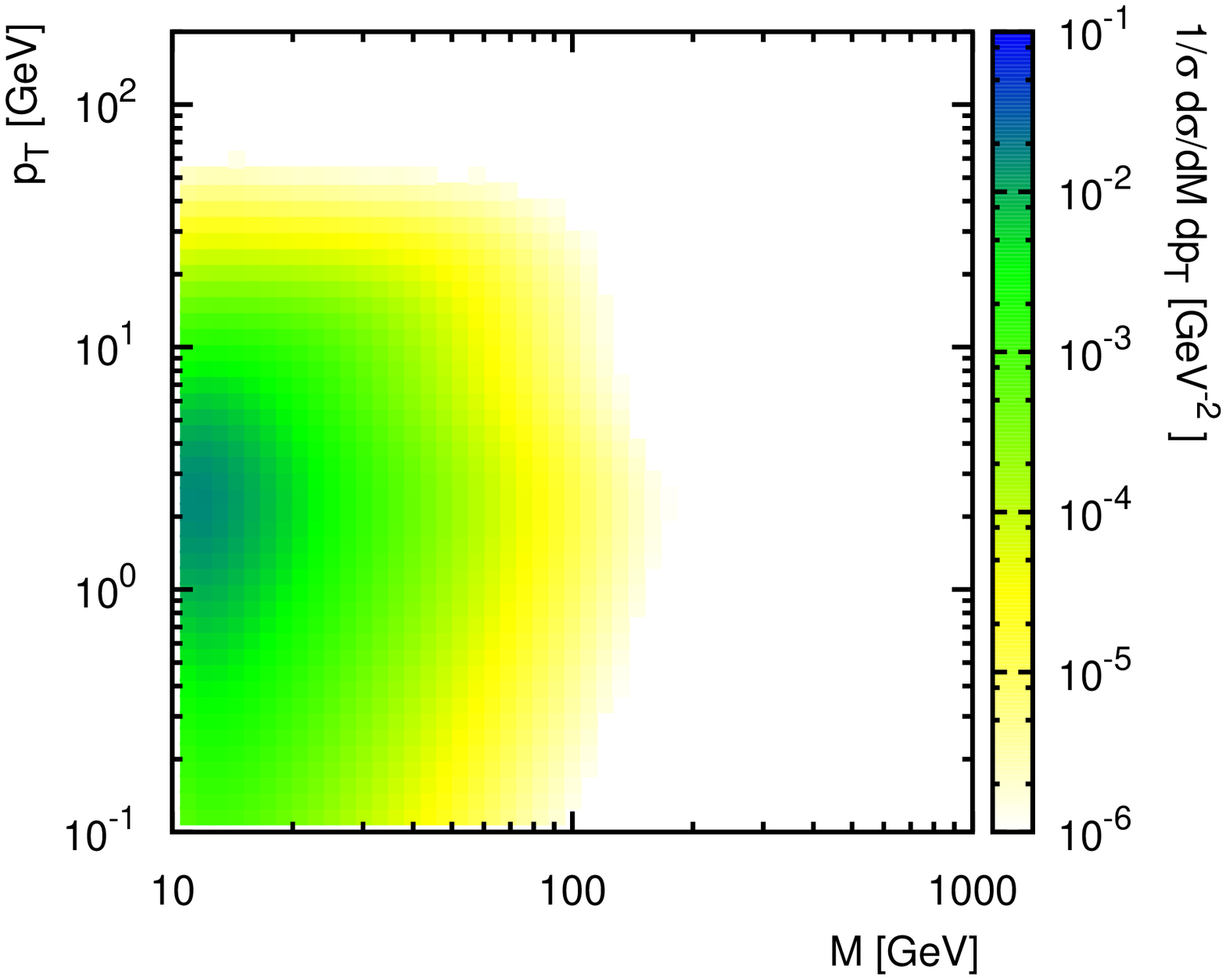, width = 5.4cm, angle = 270}
\hspace{-1cm}
\epsfig{figure=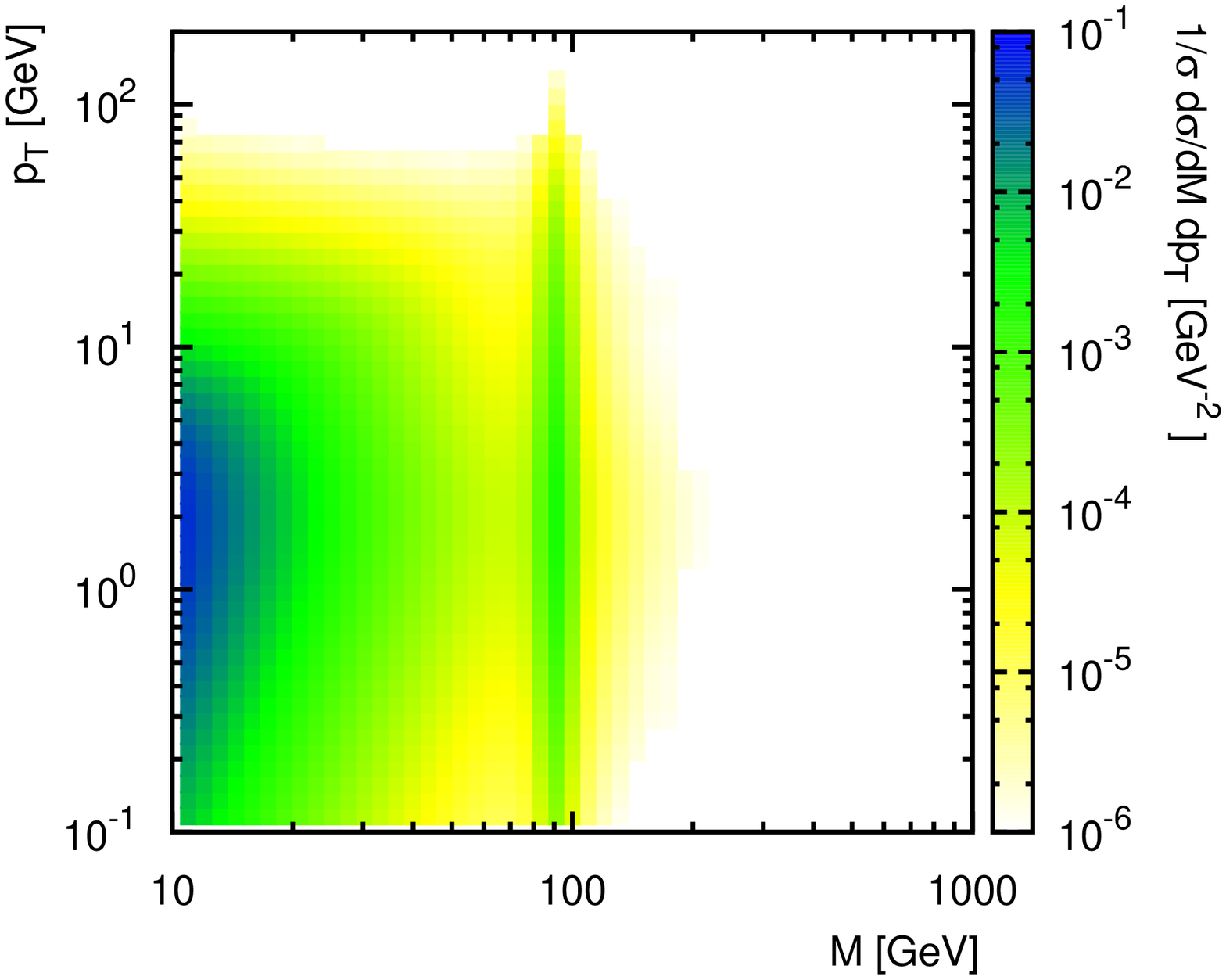, width = 5.4cm, angle = 270}
\caption{Double differential cross sections of the $b\bar b$ (left panel) and 
Drell-Yan lepton pair production (right panel)
as a functions of invariant mass $M$ and $p_T$ of the produced pair.}
\label{fig3}
\end{center}
\end{figure}

\begin{figure}
\begin{center}
\epsfig{figure=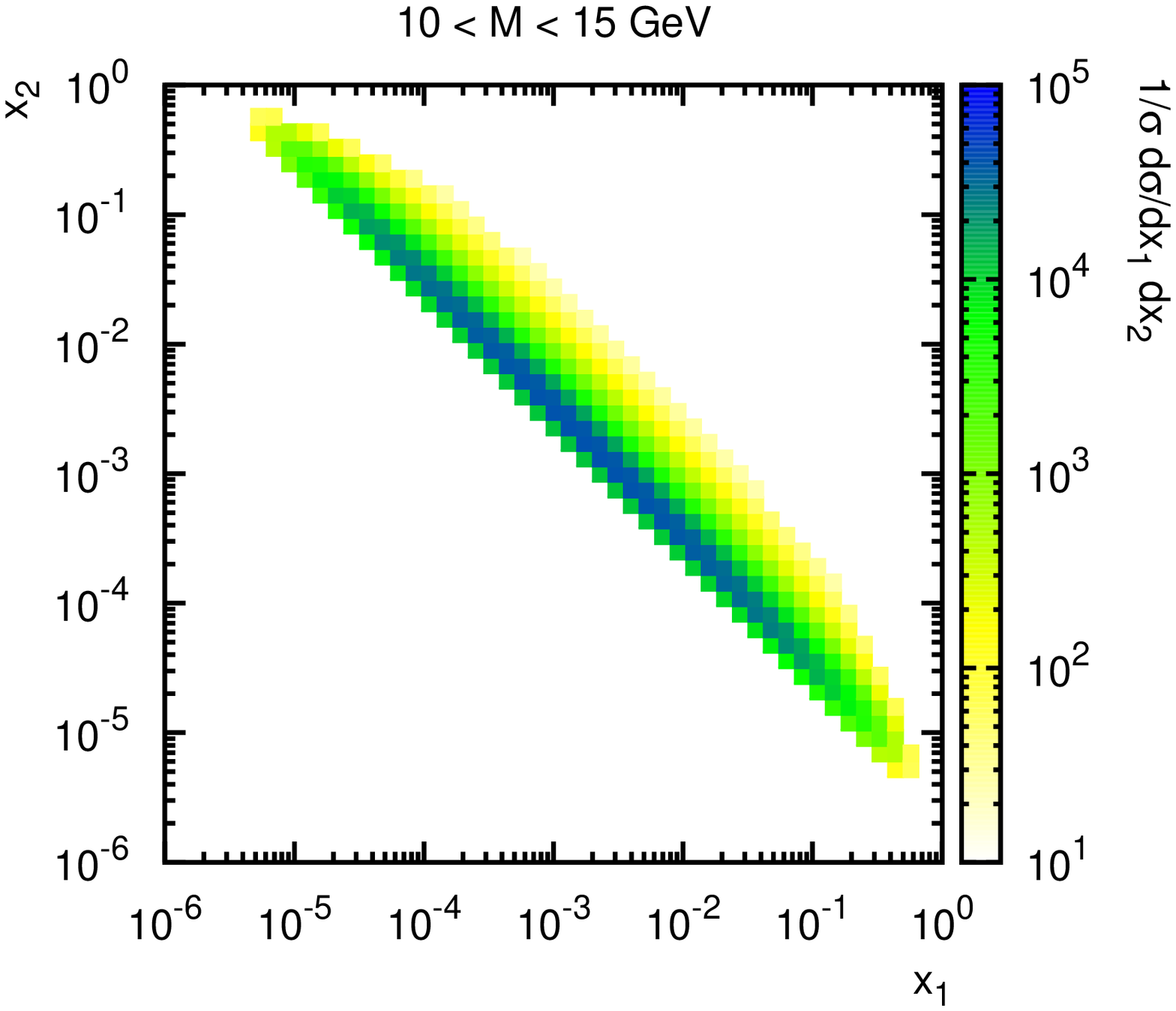, width = 5.5cm, angle = 270}
\hspace{-1cm}
\epsfig{figure=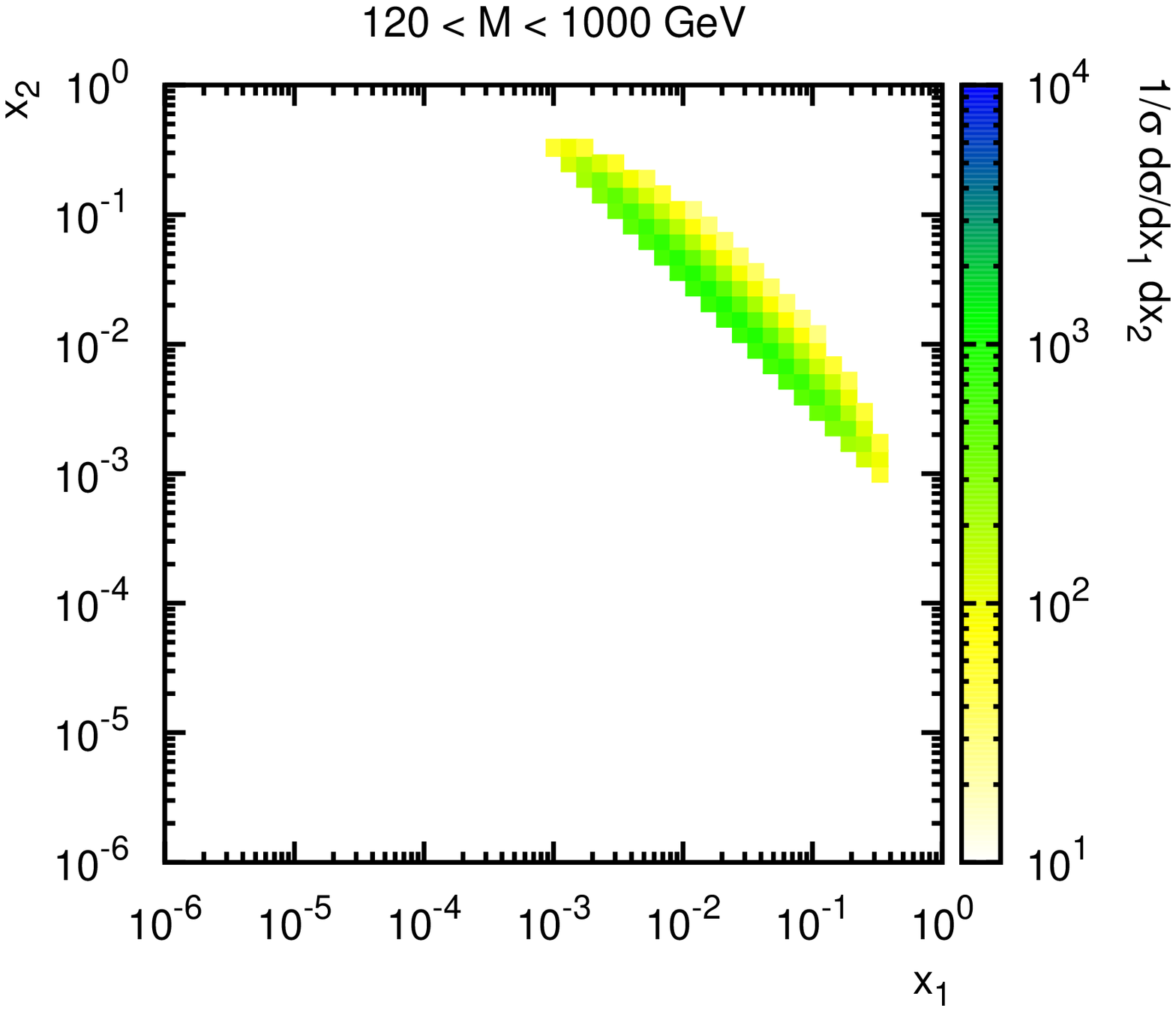, width = 5.5cm, angle = 270}
\caption{Double differential cross sections of the $b\bar b$ pair production
as a functions of $x_1$ and $x_2$ for several different intervals of the $b\bar b$ invariant mass $M$.}
\label{fig4}
\end{center}
\end{figure}

\begin{figure}
\begin{center}
\epsfig{figure=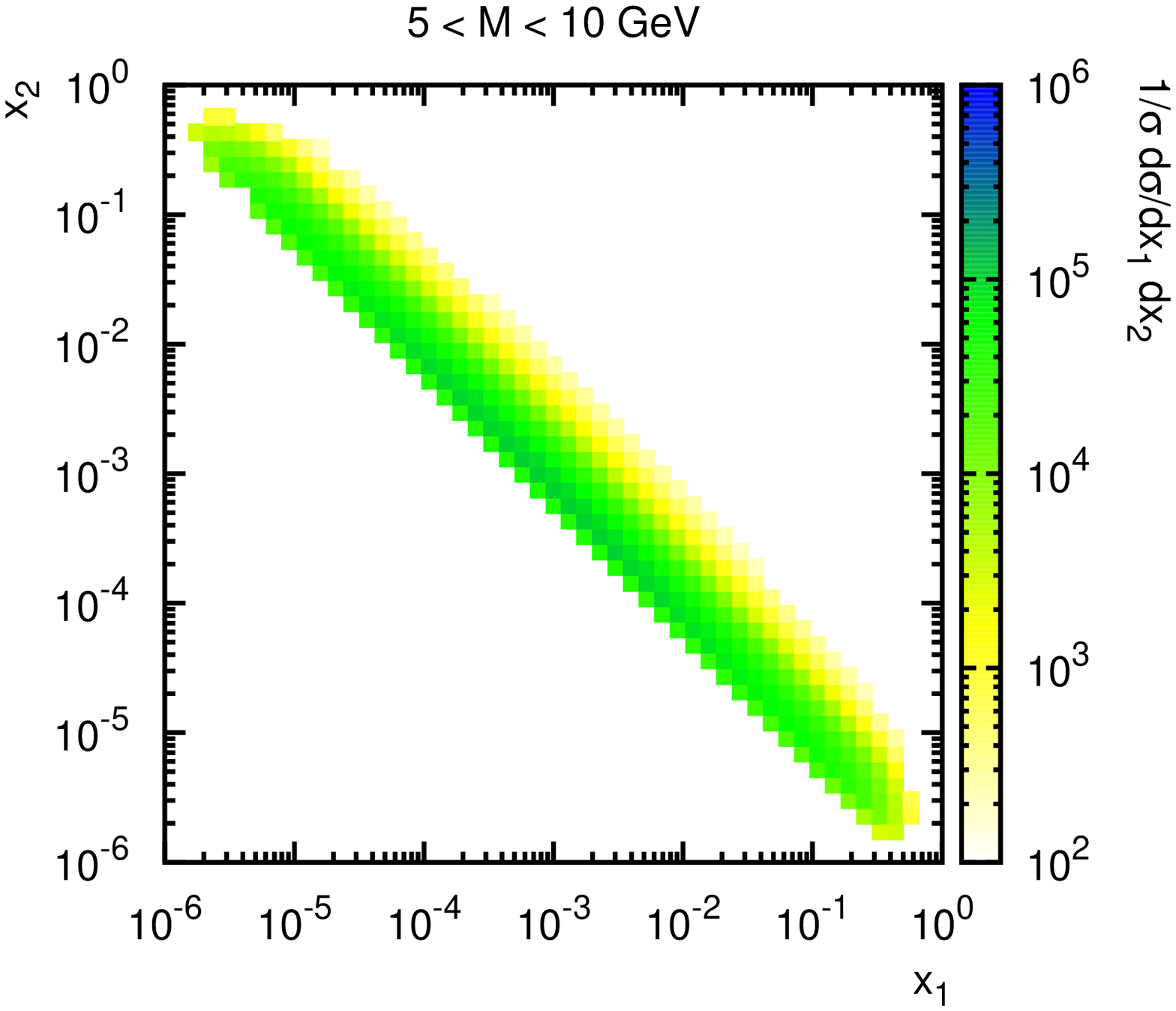, width = 5.5cm, angle = 270}
\hspace{-1cm}
\epsfig{figure=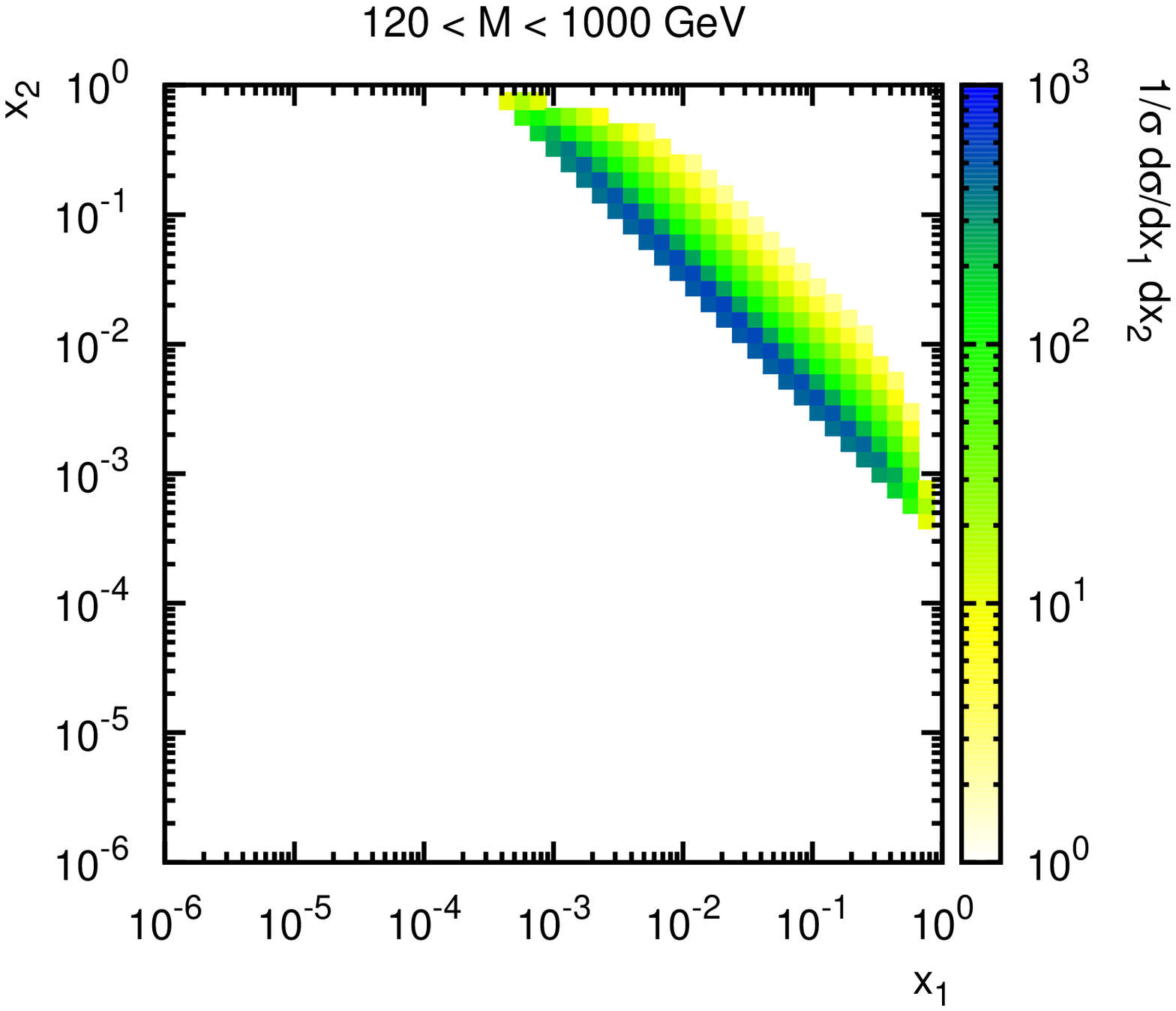, width = 5.5cm, angle = 270}
\caption{Double differential cross sections of the Drell-Yan lepton pair production
as a functions of $x_1$ and $x_2$ for several different intervals of the dilepton invariant mass $M$.}
\label{fig5}
\end{center}
\end{figure}

\begin{figure}
\begin{center}
\epsfig{figure=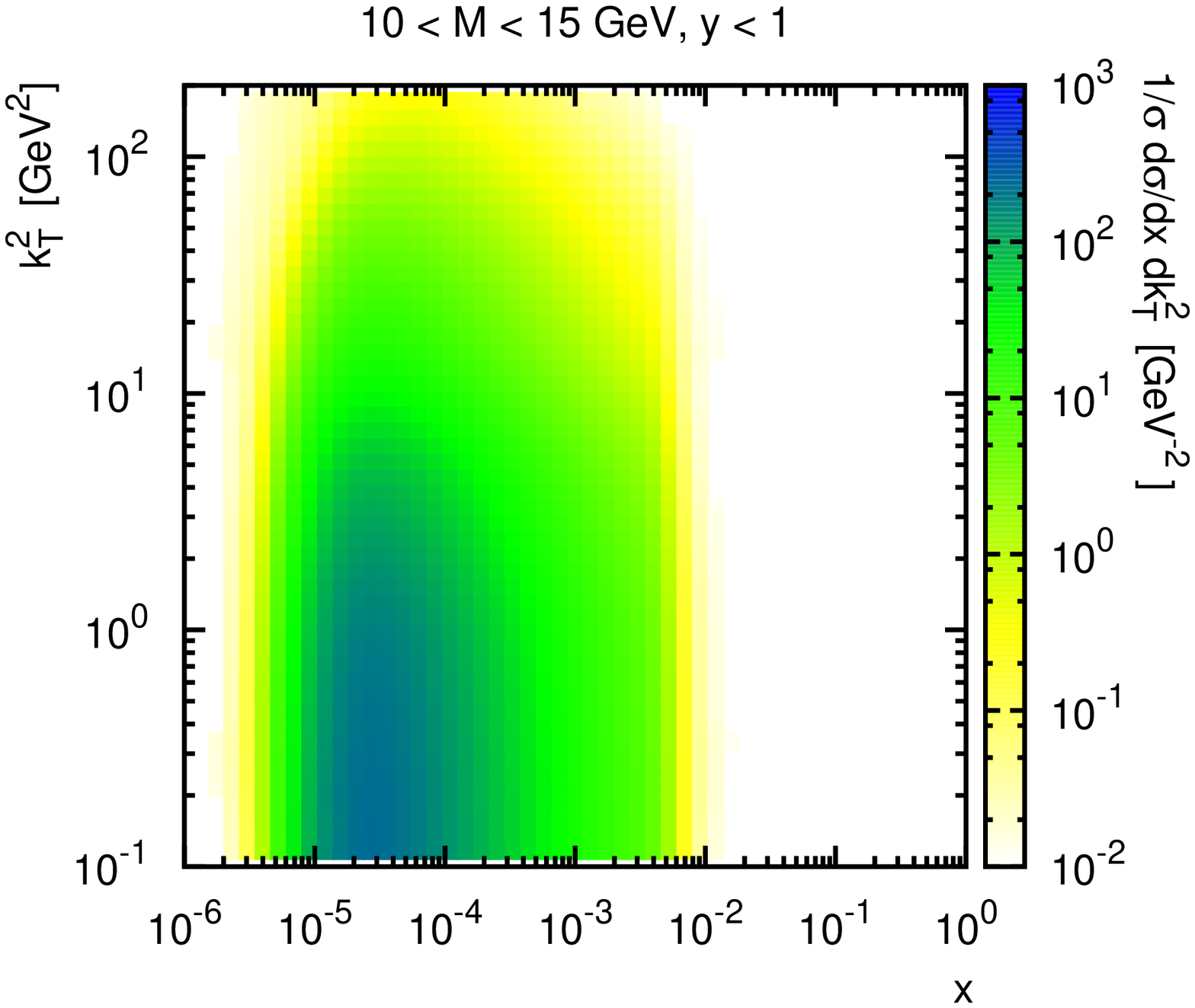, width = 5.5cm, angle = 270}
\hspace{-1cm}
\epsfig{figure=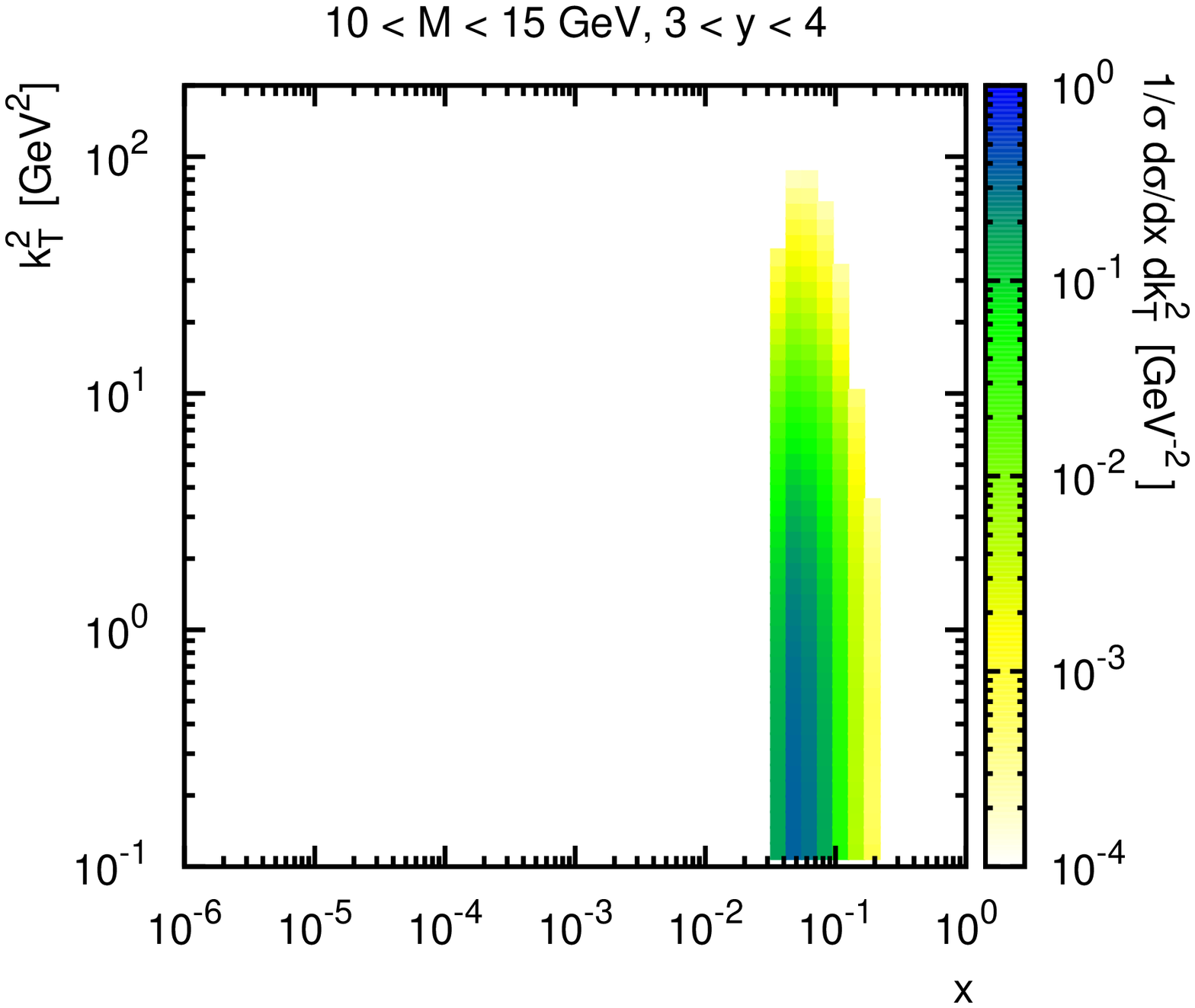, width = 5.5cm, angle = 270}
\epsfig{figure=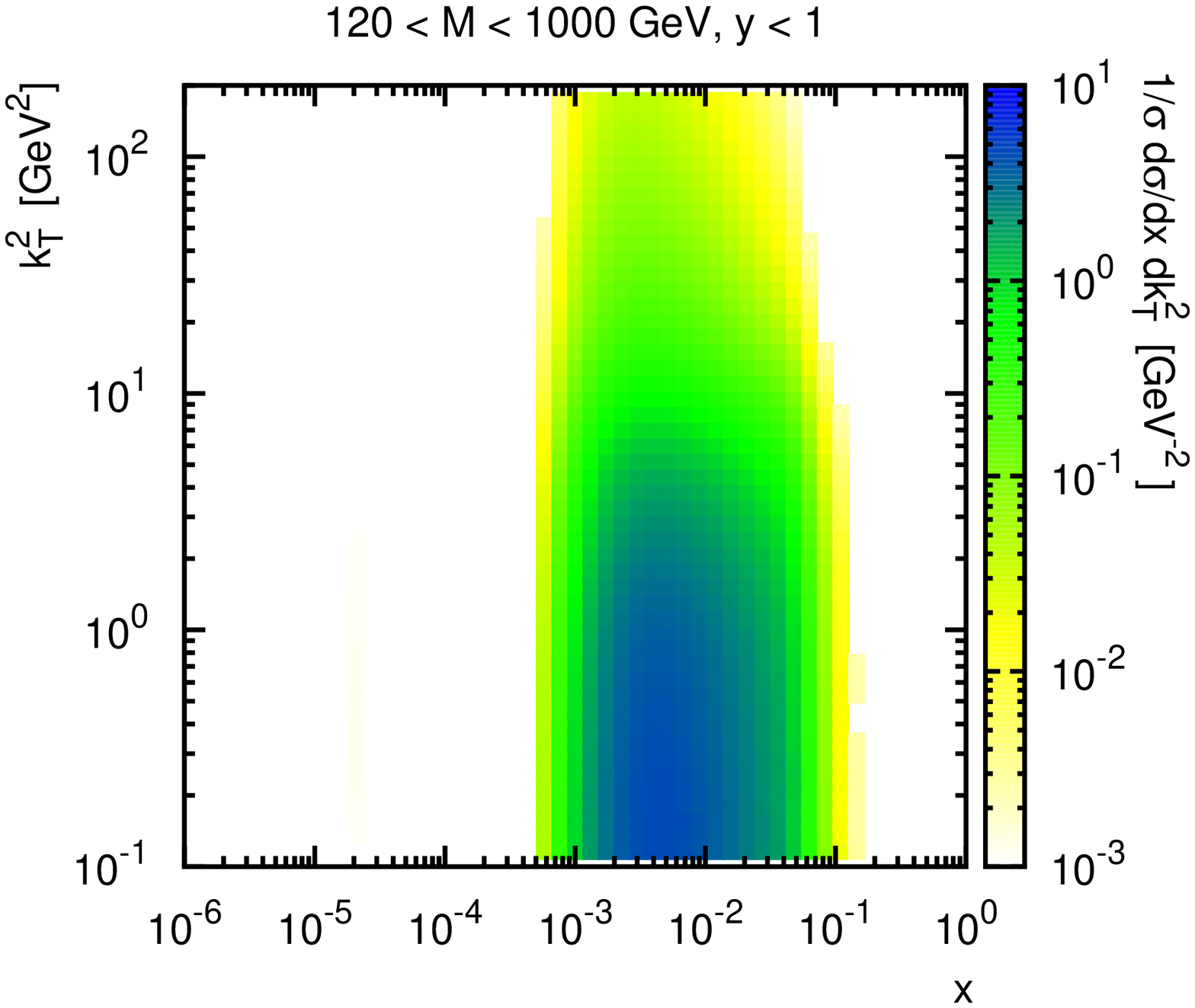, width = 5.5cm, angle = 270}
\hspace{-1cm}
\epsfig{figure=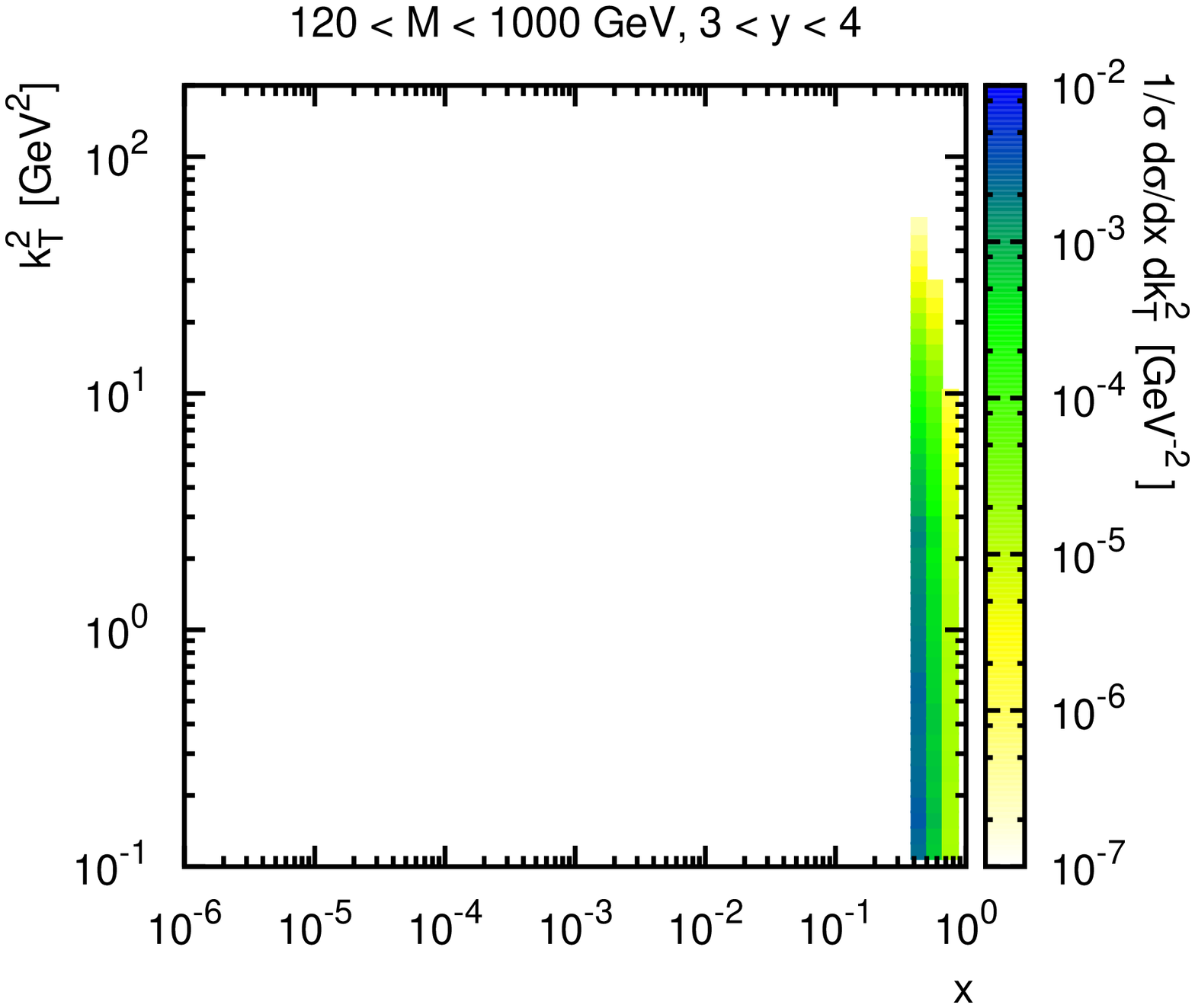, width = 5.5cm, angle = 270}
\caption{Double differential spectra of the $b \bar b$ pair production as a function of
$x$ and ${\mathbf k}_T^2$ for several different intervals of the $b\bar b$ invariant mass $M$ and rapidity $y$.}
\label{fig6}
\end{center}
\end{figure}

\begin{figure}
\begin{center}
\epsfig{figure=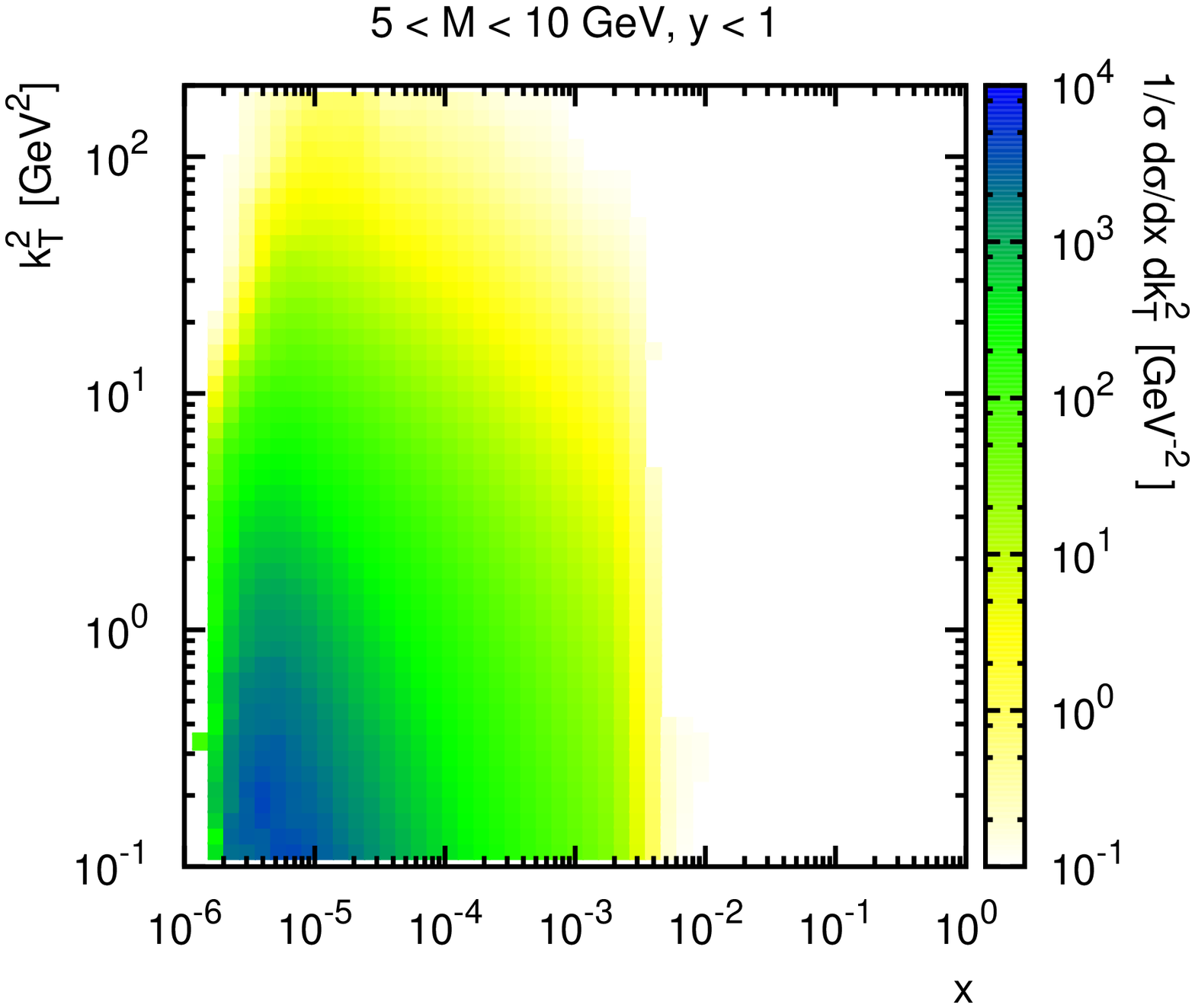, width = 5.5cm, angle = 270}
\hspace{-1cm}
\epsfig{figure=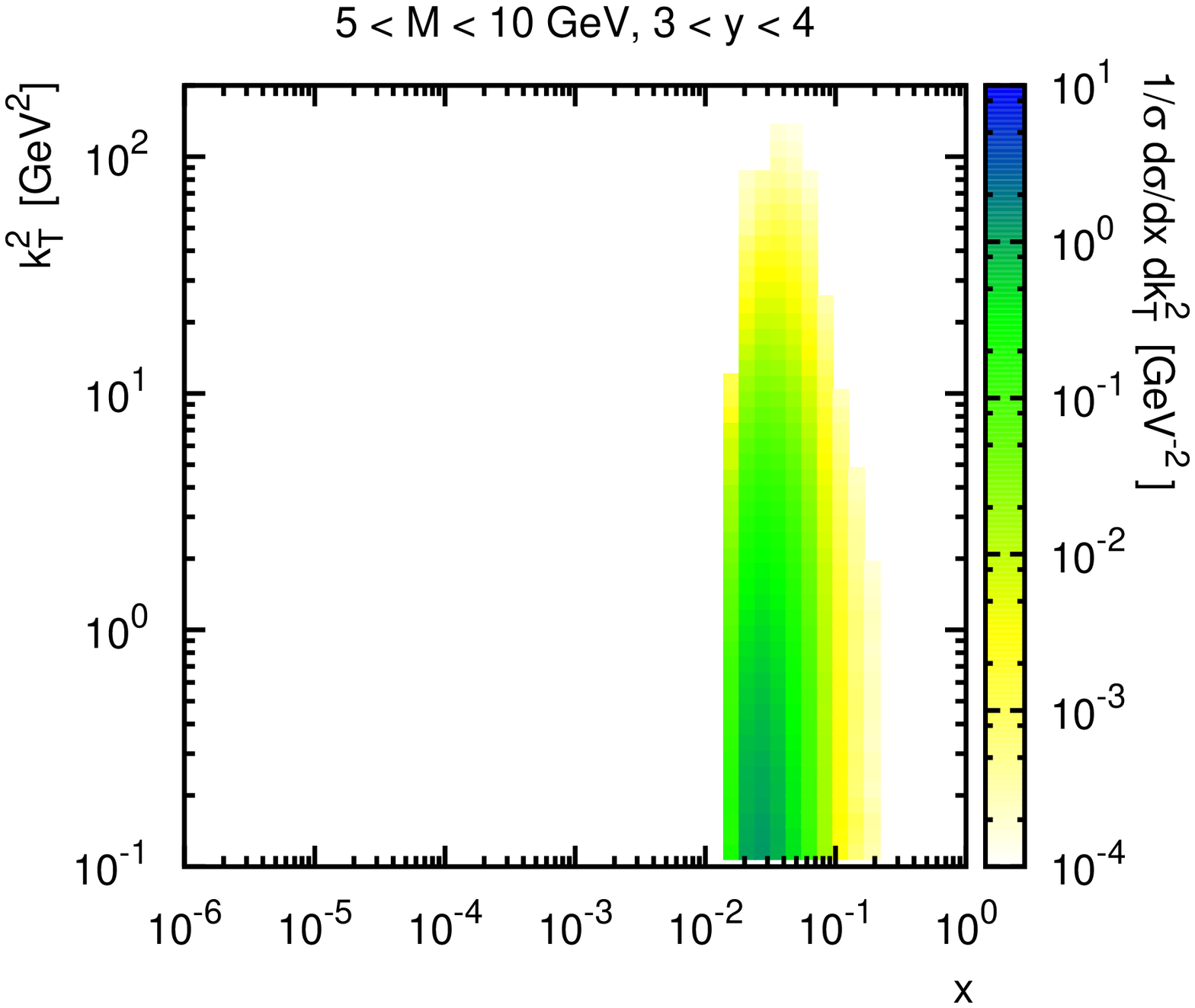, width = 5.5cm, angle = 270}
\epsfig{figure=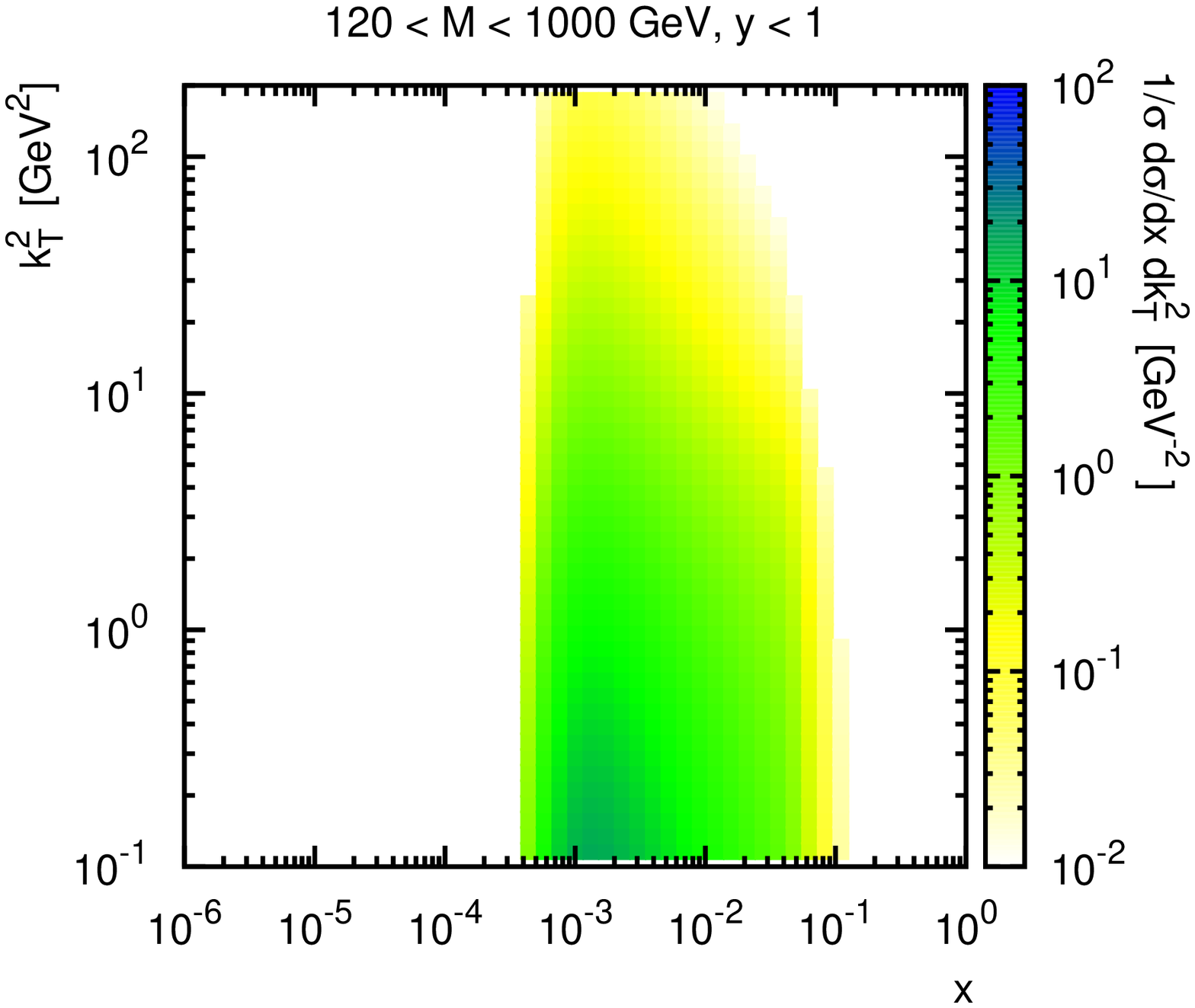, width = 5.5cm, angle = 270}
\hspace{-1cm}
\epsfig{figure=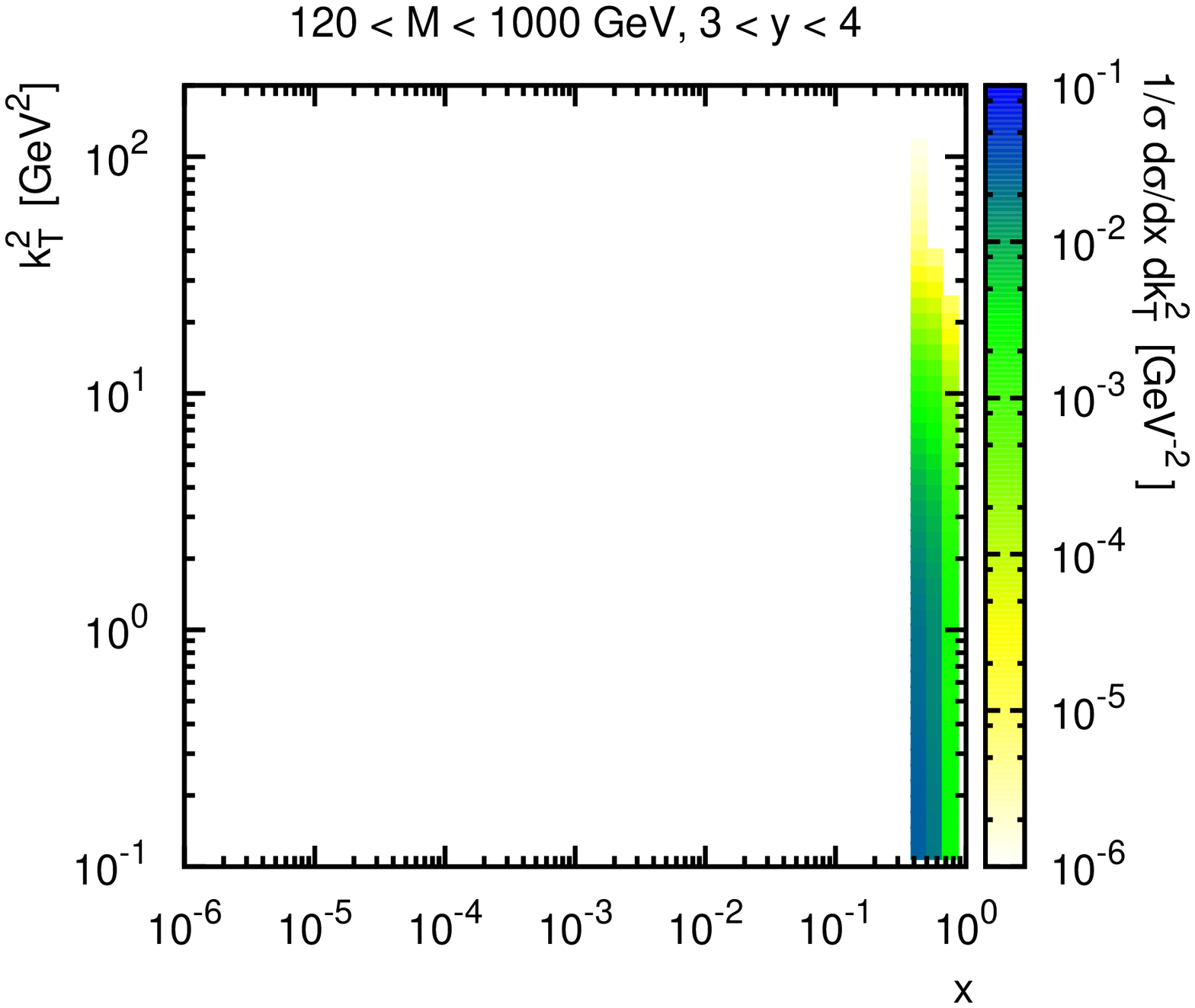, width = 5.5cm, angle = 270}
\caption{Double differential spectra of the Drell-Yan pair production as a function of
$x$ and ${\mathbf k}_T^2$ for several different intervals of the dilepton invariant mass $M$ and rapidity $y$.}
\label{fig7}
\end{center}
\end{figure}

\end{document}